\documentclass[usenatbib,usegraphicx]{mn2e}
\usepackage{rotating}
\usepackage{longtable}

\newcommand{\mum}{$\,\mu$m}

\newcommand{\wmap}{\textsl{WMAP}}
\newcommand{\planck}{\textsl{Planck}}

\newcommand{\mnras}{MNRAS}
\newcommand{\apj}{ApJ}
\newcommand{\apjl}{ApJ}
\newcommand{\aap}{A\&A}
\newcommand{\nat}{Nature}
\newcommand{\prd}{Phys.~Rev.~D}

\begin{document}
\title[Arp~220 sub-mm polarization]
{An upper limit to polarized submillimetre emission in Arp~220}

\author[Seiffert et al.]{
Michael Seiffert$^{1}$\thanks{E-mail: michael.seiffert@jpl.nasa.gov},
Colin Borys$^{2,3}$,
Douglas Scott$^{4}$,
Mark Halpern$^{4}$ \\
\\
$^{1}$ Jet Propulsion Laboratory, 4800 Oak Grove Drive, Pasadena, CA 91109,
 USA \\
$^{2}$ California Institute of Technology, Pasadena, CA 91125, USA\\
$^{3}$ Department of Astrophysics \& Astronomy, University of Toronto,
			Toronto, ON, M5S 3H4, Canada \\
$^{4}$ Department of Physics \& Astronomy,University of British Columbia,
       Vancouver, BC, V6T 1Z1, Canada
}

\date{\today}
\maketitle

\begin{abstract}
We report the results of pointed observations of the
prototypical ultra-luminous infrared galaxy (ULIRG) Arp~220 at 850\mum\ using
the polarimeter on the SCUBA instrument on the James Clerk Maxwell Telescope.
We find a Bayesian 99 per cent confidence upper limit on the polarized emission
for Arp~220 of 1.54 per cent, averaged over the 15 arcsec beam-size.
Arp~220 can
serve as a proxy for other, more distant such galaxies. This upper limit
constrains the magnetic field geometry in Arp~220 and also provides evidence
that polarized ULIRGs will not be a major contaminant for next-generation
cosmic microwave background polarization measurements.
\end{abstract}

\begin{keywords}
galaxies: individual: Arp~220 -- galaxies: magnetic fields
-- cosmic microwave background
\vspace*{-1.25cm}
\end{keywords}

\section{Introduction}

The new generation of Cosmic Microwave Background (CMB) experiments,
such as NASA's Wilkinson Microwave Anisotropy Probe (\wmap) and a number of
balloon-borne and ground-based instruments, are revolutionizing
cosmology by providing precision estimates of a number of fundamental
cosmological parameters \citep[e.g.][]{bennett03, tegmark04, mactavish05, spergel06}.
The upcoming launch of ESA's \planck\ mission
\citep{bersanelli96, bluebook05}
will provide an opportunity for even more precise parameter estimates.
 
In addition to the temperature anisotropy, the CMB is expected to be 
partially polarized due to Thomson scattering of the anisotropic
radiation field near the surface of last scattering.
The first definitive detections of CMB polarization have recently been
made \citep{dasi,wmap,cbi,capmap,boomerang,page06}.
The angular power spectrum of polarized fluctuations
can provide a wealth of additional cosmological information
\citep[see e.g.][]{huandwhite}. Perhaps the most tantalizing prospect
is that primordial gravitational waves from the epoch of inflation will
leave a distinct divergence-free (or `$B$-mode') signature in the CMB polarization
that may be detectable by future experiments and cleanly separated from the curl-free (or
`$E$-mode') dominant polarization signal \citep{zaldarriaga97,
kamionkowski97}.

To realize this potential, careful control of systematic effects,
including foreground emission, is essential.  A number of 
studies have characterized the potential of extragalactic sources
to contaminate CMB anisotropy measurements
\citep[e.g.][]{toffolatti98,tegmark00,tucci04,toffolatti05}.
In general, these authors conclude that extragalactic contamination
is most important at small angular scales (high multipole moments),
that current estimates are not precisely constrained by available 
measurements, and that these sources are unlikely to pose a major
difficulty to future CMB measurements. 

 \citet{scott99} analyzed the
data from early SCUBA surveys at 850\mum\
\citep[e.g.][]{smail97,hughes98, barger98, eales99}
and concluded that the \planck\ mission may be 
confusion limited at frequencies of $350\,$GHz and higher, and that clustering
of faint sources may be a measurable signal.

\citet{boryscmb} provided the first limit of the contribution of
SCUBA point sources to the CMB.  Several studies
\citep[e.g.][]{haiman,gonzalez05} have modeled 
how clustering of such sources may be manifest in the background 
over a wide range of wavelengths.

The potential of polarized extragalactic sources to contaminate
polarized CMB anisotropies has been much less well studied, 
particularly in the sub-mm region \citep{dezotti99,tucci04}.  In this paper, 
we address this concern by investigating polarized sub-mm emission 
from one particular object.

Arp~220 is a prototypical member of the class of ultra-luminous
infrared galaxies (ULIRGs), with a far-IR luminosity of approximately
$1.6 \times 10^{12} L_\odot$ \citep{soifer84,lisenfeld00}. 
At roughly $70\,$Mpc ($H_0=70\,{\rm km}\,{\rm sec}^{-1}{\rm Mpc}^{-1}$)
distance, it is the closest member of this class and one of the
brightest galaxies in the local Universe. It is therefore well
suited for studies that would be much more difficult for other,
more distant ULIRGs. 

Although our initial motivation for investigating the sub-mm polarization
of Arp~220 was as a means to estimate CMB contamination, such measurements
of polarization are interesting in their own right.
Studying regions of polarized dust emission is 
important as a means of probing the magnetic field geometry
responsible for aligning the dust grains \citep{lazarian03}.
Related issues include understanding the source of sub-mm emission,
galactic superwinds, internal dynamics and dust physics.

\section{Observations}
We observed Arp~220 with the Submillimetre Common User Bolometer
Array (SCUBA, Holland et al.~1999) at the James Clerk Maxwell 
Telescope (JCMT), Mauna Kea on 2000 August 23 and again on 2001 March 16.
Conditions in the 2000 run were favourable, with an 850\mum\ 
opacity of $\sim 0.3$  ($\tau_{\rm CSO}\sim 0.07$). The sky was much 
less opaque in the 2001 run, where we enjoyed roughly a factor 
of 2 lower opacity. Observations were conducted using the SCUBA polarimeter
\citep{scubapol} which consists of a rotating quartz half-wave plate in
front of a fixed wire grid analyzer, mounted externally on the SCUBA
dewar.  

We chose to perform the polarization observations in `photometry' mode as
opposed to the more common imaging mode, in order to achieve higher on-source
efficiency.  The angular size of Arp~220's infrared luminous core is smaller
than the resolution of SCUBA at even its highest frequency channel, and thus
we are not missing any flux by performing a photometry-mode observation.
Because SCUBA employs a dichroic beam splitter, data are collected 
simultaneously for the 450\mum\ and 850\mum\ arrays.  Although we are mainly
interested in the central array bolometers (denoted `C14' and `H7', for the
short and long wavelength arrays, respectively),  data from the remaining array
bolometers were also gathered and used as a monitor of the atmospheric emission.

The observations required several levels of signal modulation. 
The array was chopped in azimuth at $7.8125\,$Hz in order to remove 
common mode atmospheric signal in the source and reference positions.
The chop throw was 90 arc seconds for the 2000 observations 
and 40 arc seconds for the 2001 observations. 
After  four, 1 second integrations, the telescope was `nodded' in 
azimuth to match the fast azimuth chop and another set of four, 1 second 
integrations were taken.  The difference between the chopped signal in 
both nods removes instrumental effects which are dependent on the secondary
mirror position \citep{zemcov05}.

After these integrations, the polarimeter half-wave plate was moved to the
next in a sequence of 16 rotational positions.  A full set of rotational 
positions was thus obtained, consisting of 128 seconds of observations, 
with an elapsed time of 280 seconds, including overheads.  Arp~220 was 
observed at elevation angles ranging from approximately 30 to 85 degrees 
above the horizon. From the two runs, the total on-source observing time 
was approximately 3.4 hours, consisting of 95 full wave plate cycles.  
As we explain later, however, some data were not included in the analysis.

In the 2000 observing run, several levels of systematic controls were 
performed. Two different sequences of wave plate positions were used:
the first sequence was the standard one, with sixteen $22.5^\circ$ rotational
steps, one after another; in the second method, the angular sequence was 
$0^\circ$--$315^\circ$ in steps of $45^\circ$, and then 
$22.5^\circ$--$337.5^\circ$ using the same step. Observations
with the two sequences were interleaved. The different wave-plate strategies 
were performed in an attempt to detect or limit the contribution of atmospheric
fluctuations to the polarization signal.
Additional observations were performed in a similar manner using
a different 850\mum\ bolometer (`G15') centred on Arp~220.
After reducing these data and not finding any significant difference between
the different approaches, we decided that for the second run the more
straightforward approach of
using the default wave-plate position order and H7 as the primary bolometer
was better suited for observations of Arp~220.

A polarization `standard', DR 21 was also observed for 15 total
wave-plate cycles. This galactic
star-forming region has previously been observed to be 
bright and polarized. \cite{greaves99} found $2.34\pm0.27$ per cent
polarization at 800\mum\ with a position angle of $20^\circ \pm 3^\circ$.
\citet{minchin94} report  $1.8\pm0.3$ per cent polarization at
a position angle of $17^\circ \pm 4^\circ$, also at 800\mum.
DR21's relatively high flux and likely lack of variability make this
source convenient for cross-checking polarization measurements.

Absolute flux calibration and determination of the instrumental
polarization (IP) were provided by observations of Uranus 
(13 wave-plate cycles) and Mars (12 wave-plate cycles).
The Uranus observations were conducted with the same variety
of systematic checks as the Arp~220 observations. 
Pointing was checked roughly once per hour using the blazar 1611+343. 
Pointing corrections were typically a few arcseconds, except near 
transit -- because the JCMT is an alt-az telescope, it has difficulty
tracking sources such as Arp~220 that transit near the zenith. 
Arp~220 reaches an elevation of $\sim85^\circ$ at the latitude of
the JCMT.

\section{Data Analysis and Results}
The raw data were reduced with a combination of routines
from the SCUBA User Reduction Facility \citep[SURF][]{surf}
and our own software. {\sc SURF} was used to provide 
data free from atmospheric signal, 
while our own code 
was devoted to estimating the polarization strength. The roles of the
codes are described below.

\subsection{Preliminary processing}
Using {\sc SURF}, the data from the two nod positions were subtracted,
flat-fielded and then corrected for extinction.  We used the
polynomial fits to the CSO 225\,GHz opacity from the JCMT web 
page\footnote{See {\tt http://www.jach.hawaii.edu} for more information.}.
The opacity at 225\,GHz was then converted to the 450 and 850\mum\ 
bands using the relations in \citet{archibald}. 
The residual sky background was subtracted \citep{jenness98}, as estimated
from the median of the twelve lowest noise bolometers on ring 3 of the array, 
and additionally a few anomalous 5-$\sigma$ or greater spikes were
removed. 

\begin{figure*}
\begin{center}

\includegraphics[width=1.0\textwidth,angle=0]{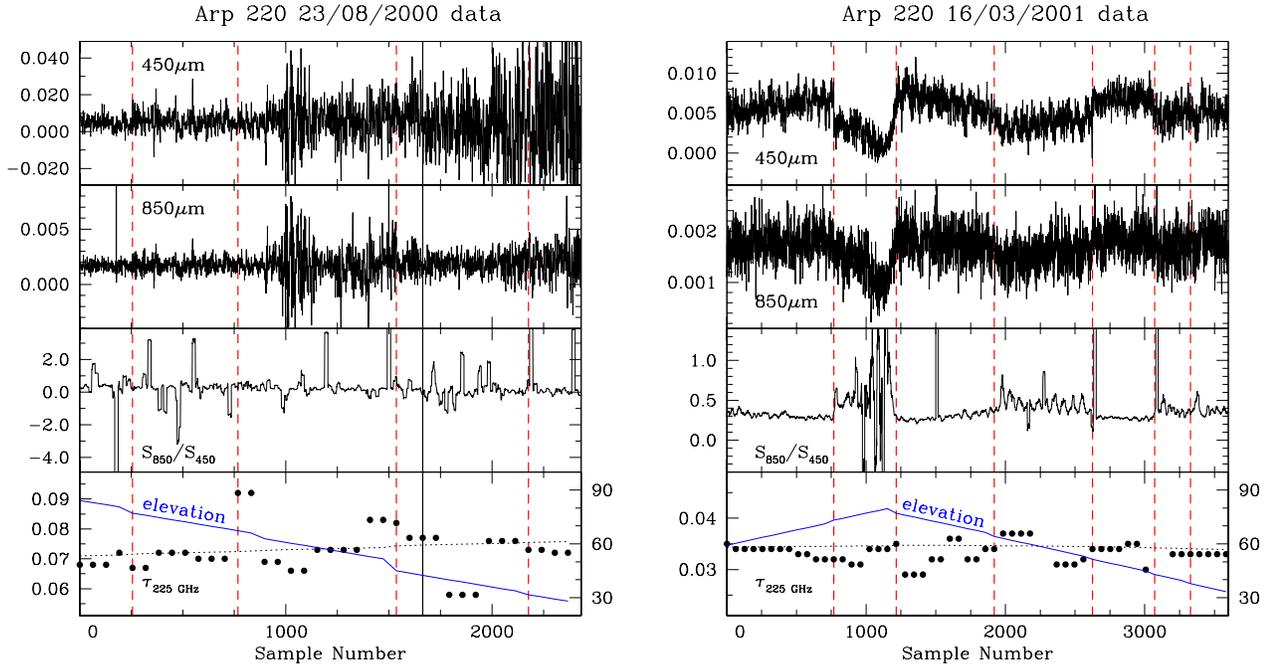}
\caption{
Timestreams for the extinction corrected, sky-subtracted 
on-source bolometer from the two runs.
The top two panels show the 450 and 850\mum\ signal (in Volts). The 
third panel shows the ratio of the 2 signals, and should be constant
under ideal conditions. The bottom plot shows the measured 225 GHz CSO $\tau$
(black circles)  and the polynomial fit (dotted line), together with the
elevation angle of the observation (solid line).
Vertical dashed lines indicate where pointing checks were performed.
{\it Left\/}: The 2000 data are shown, with a solid black vertical line at 
sample 1664, denoting the transition from
observations taken with H7 as the on-source bolometer to using bolometer G15.
{\it Right\/}: The 2001 data highlights the problems associated with observing
the target near the zenith. In particular,
the data between samples ${\sim}\,750$ and  ${\sim}\,1250$ demonstrate that
the source probably drifted away from the bolometer centre around transit.
These data were not included in the analysis.  The opacity was much lower
in the 2001 observations, and one can actually see the instrumental
polarization in the 450\mum\ data.
}
\label{fig:timestreams}
\end{center}
\end{figure*}

Since we expect that the polarization will be weak, particular care
was taken to ensure the quality of the data that went into the analysis.
In Fig.~\ref{fig:timestreams} we plot the time-streams of the central
bolometer in each array for both runs.

\subsection{Systematic error control}
Among the many sources of potential systematic error for 
polarization measurements, atmospheric transmission fluctuations,
pointing errors or drifts, and spurious pick-up from the telescope
environment are perhaps the most important \citep{hildebrand00}.

Contaminating pick-up was evaluated by processing the data from an
off-source bolometer through the entire analysis pipeline. We have done
this for the 'H9' bolometer. The polarized signal from this bolometer was
consistent with zero, at the 1-$\sigma$ level, and had a similar noise level to
the on-source bolometer analysis. We therefore conclude that polarized
contamination
from the ground or from atmospheric emission (which we would expect to 
contaminate the on- and off-source bolometers at a similar level) does not
contribute significantly to our on-source observations. 

We have examined the time series data for evidence of atmospheric transmission
fluctuations.  These would affect an observation of a bright
point source differently than an off-source pixel, and may be expected to
add noise to the measurement. If such fluctuations were significant and
had somewhat
shorter time-scale than the wave-plate rotation period, they would introduce a 
spurious signal that could be confused with intrinsic source polarization.
One may also expect that a decrease in
atmospheric transmission would be correlated with an increase in atmospheric
emission (if due to a short time-scale fluctuation in opacity). 
In examining the time series data for the on- and off-source
bolometers for our observations, we have found no evidence for such an effect.
We are only sensitive to the spatial gradients in off-source emission, however,
because only the differenced data after the fast chop are available from SCUBA.

By examining the full time series of both the 450 and 850\mum\ data, we have
concluded that some observations were drifting somewhat off-source.
A key signature of such a drift is a significant decrease in the amplitude of 
the observed signal, and a greater change at 450\mum\ than at 850\mum\ due to
the differences in beam-size. Such a decrease is evident near sample number 1000
in the right hand panel of Fig.~1.  We have excluded from our final analysis
the 7 full wave-plate cycles of data 
between pointing checks near where this pointing drift occurred.
We note that the telescope was pointed at very high elevation angle
during this period, where tracking is most difficult.  A comparison with and
without these data included shows that our final polarization result changes by
less than 1-$\sigma$.  Very small pointing drifts that may exist in the
remaining data are not expected to affect the final result, since
each wave-plate cycle has a gradient removed from the fit to $Q$ and $U$.

A period of increased noise near sample number 1000 in the left hand panel of
Fig.~1 is evident. A comparison of data before and after sky subtraction
shows that this is due to a period of increased sky noise.  Excluding
this section of data make less than a 0.3-$\sigma$ change in the Arp~220
polarization
fraction, and this section of data has been retained in our final result.
In order to test the robustness of our results we have experimented with
removing additional short sections of data.  Aside from the 7 wave-plate cycles
mentioned above,
we have not found the results to be sensitive to the removal of other sections,
beyond the expected increase in overall noise.

Our observations of DR21 formally detect 850~$\mu$m polarization at
approximately the 4-$\sigma$ level and are consistent (at the 1-$\sigma$ level)
with those reported by \citet{minchin94}.  It is not clear to us, however,
that our off-source chop or sky subtraction is always in an emission-free
region, so our claim for DR21 is one of broad consistency with previous
measurements, rather than of precise determination.

\subsection{Polarization analysis}
Using our own software, the average signal and estimated error for the four
integrations at each wave-plate position for a full wave-plate
cycle were fit (using singular value decomposition) to provide an estimate of the total
flux and the degree and direction of polarization. A linear gradient 
in flux was removed from each waveplate cycle in an attempt to
decrease the effect of residual pointing drifts. 
The results for all the wave-plate cycles were then combined after discarding
several waveplate cycles due to pointing drift as described above. 
This fitting procedure essentially
mirrors that of the observatory's available software, but allowed us to pipeline
the analysis and rapidly compare a number of alternative reduction schemes before
concluding that the processing described here was sufficient. Among the alternatives
investigated (and eventually passed over) were fits to partial and multiple waveplate rotations, discarding waveplate rotation cycles based on a goodness-of-fit criterion,
discarding waveplate rotations that produced outliers in $Q$ and $U$, and an attempt at regressing fluctuations as measured with the 450~$\mu$m data.
An overall instrumental polarization (IP) that is contributed
mostly by the JCMT wind screen must also be subtracted before
arriving at the final source polarization estimate.
We removed the IP estimated from previous observations of Mars and Uranus
(assumed to be unpolarized) of 1.06 per cent at a position angle of
$161^\circ$ given by \citet{scubapol}.
Through our observations of Mars, we derive our own estimate
of the IP of $0.89 \pm 0.23$ per cent at $154^\circ \pm 7^\circ$, which is
consistent with the \citet{scubapol} value.  Our final results are
relatively insensitive to the precise value of the IP, because of
the range of elevation angles during which the Arp~220 measurements
were made. Switching between the two estimates of IP above gives
less than a 1-$\sigma$ change in the final results for Stokes $q$ and $u$.
We use the values from \citet{scubapol} for our final results.

The data reduction process was run on all the planetary observations to 
produce an estimate of the system calibration. 
We adopt a value of $425\,{\rm Jy}\,{\rm V}^{-1}$ for the results presented
here. There is a systematic uncertainty in this value from atmospheric
and pointing effects, which we estimate to be approximately $\pm$5 per cent
based on the distribution of the calibration data.  An overall 
change in the calibration, however, will not affect the polarization estimate.

The results from the first and second
observing runs are consistent, with most of the 
statistical weight coming from the second run.
Our formal statistical result for Arp~220 at 850\mum\ is 
\begin{eqnarray}
I & = & 751 \pm 38~{\rm mJy} \nonumber \\
q & = & 0.00582 \pm 0.00350 \nonumber \\
u & = & 0.00426 \pm 0.00448, \nonumber
\end{eqnarray}
where $q$ and $u$ are the normalized Stokes parameters, and $I$ is the
Stokes intensity.  Our intensity measurement is consistent with the values
measured by \citet{lisenfeld00} and \citet{slugs}.
A naive estimate for the polarization amplitude and orientation
can be produced from these
measurements, using $p^2=q^2+u^2$ and $\tan(2\theta)=q/u$.  This
yields a polarization fraction of $p=0.72 \pm 0.39$ per cent
at a position angle of $18.1 \pm 15.4$~degrees.
This value, however, considerably overestimates the true level of
polarization due to the biasing effect of noise
\citep[see][]{rice47, Ser58, WarKro, simmons85, hildebrand00}.
To counter the effect of noise bias, we can produce an estimate of the 
posterior probability distribution 
using a Bayesian framework and an assumed uniform prior on the degree
of polarization and the position angle. 
This procedure yields no firm polarization detection and 
gives our final result for the 99 per cent confidence upper
limit on the polarization of Arp~220: 1.54 per cent.

We note that pointing control is a likely systematic limitation to
these or similar measurements. A systematic wander of 2.5 arcsec
during the 12 minutes it takes to complete a wave-plate rotation is
sufficient to induce a 1 per cent polarized signal.  
These concerns are more serious for the 450\,\mum\ measurements, which have
a smaller beam-size. We therefore do not report a polarization result for the 
450\,\mum\ Arp~220 measurements.  The coming SCUBA-2 upgrade
\citep{scuba2}
will minimize the effects of pointing drift and atmospheric fluctuations
with a filled detector array and a much faster wave-plate rotation speed. 

\begin{figure*}
\begin{center}
\includegraphics[width=1.0\textwidth,angle=0]{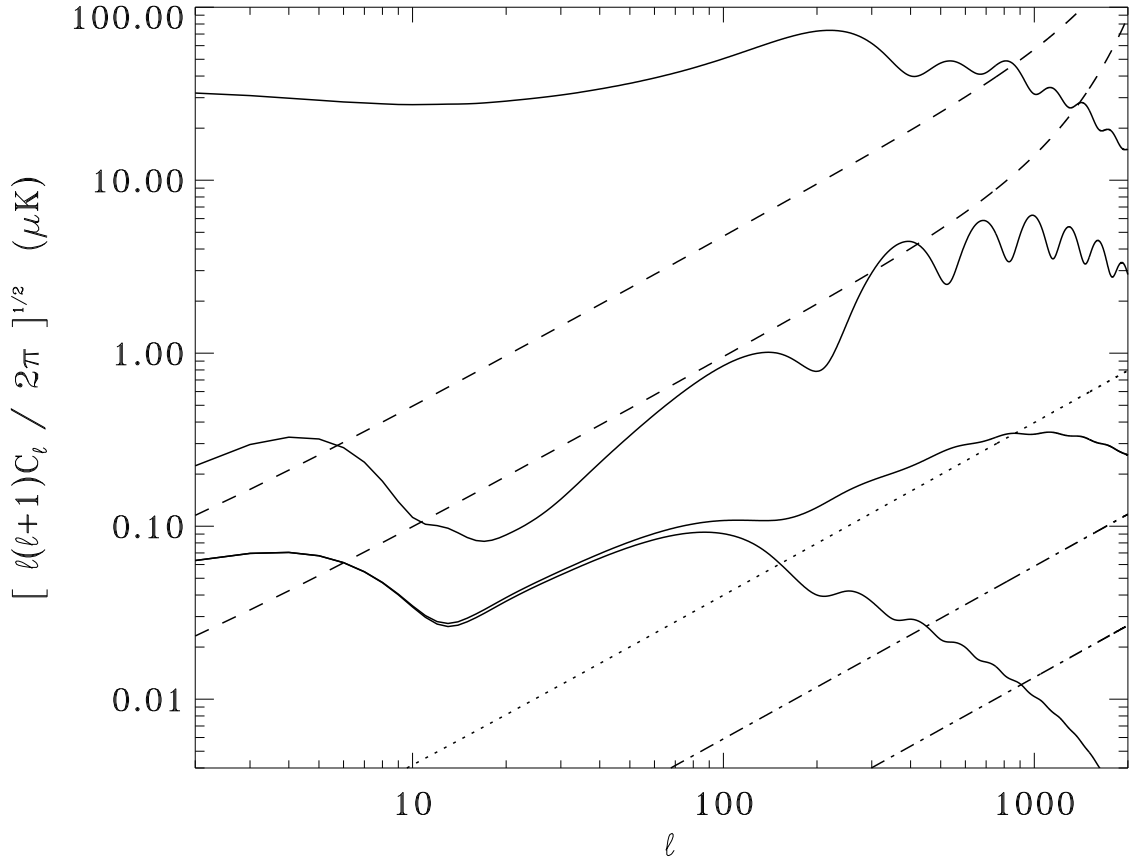}
\caption{
Theoretical CMB angular power spectra with potential contributions from
polarized dusty galaxies and with noise estimates from \planck.  The solid lines show the expected
CMB power spectrum from (top to bottom) temperature fluctuations, $E$-mode polarization, $B$-mode 
polarization with lensing, and $B$-mode without lensing. A standard cosmological model constrained
by \wmap\ three year data has been assumed, along with a tensor-to-scalar ratio of 0.1.  Also shown
are the pixel polarized noise estimates from \planck\ at 353 GHz (upper dashed line) and 143 GHz (lower
dashed line).
The dotted line shows an estimate of the contribution of polarized dusty galaxies at 350 GHz 
assuming all the sources are 1.5\% polarized with random orientation angles, the sources are not clustered,
and that sources above 100 mJy have been removed.
The two dot dash curves
show two estimates from de Zotti et al. (1999) of the contribution of dusty polarized galaxies at 143 GHz, assuming 2\% galaxy polarization.
}
\label{fig:cl_plot}
\end{center}
\end{figure*}

\section{Discussion}

The lack of sub-mm polarization signal in Arp~220 is intriguing,
but perhaps not very surprising. 
In our own Galaxy, polarized dust emission has been observed in
a variety of molecular clouds \citep[e.g.][]{fiege2000,matthews2002,houde2004}
and also in pre-stellar cores \citep{wt00}. 
Magnetic fields in these environments 
can result in aligned dust grains which emit radiation
with an E-field preferentially aligned.
The expected level of polarization is typically
a few per cent \citep{hildebrand95}.
For extragalactic sources \citet{greaves00} have reported a detection
of polarized 850\mum\ emission from M82, via resolved imaging observations with
SCUBA.  Because of the dust grain alignment mechanism, 
sub-mm polarized emission traces the magnetic field geometry
averaged over the beam-size.
Arp~220's small angular size means that the magnetic field would have to
be aligned over a significant fraction of the source in order to produce
detectable
polarization, otherwise the random orientations from a variety of distinct 
regions sampled by the JCMT beam would tend to cancel out.
Note that the average polarization over the entire M82 SCUBA mapping of
M82 carried out by \citet{greaves00} corresponds to only about 0.4 per cent.
For Arp~220 high resolution interferometric maps in CO~(1--0) and dust
continuum \citep{scoville91} show that the emission is extremely concentrated
in a dense core.  Perhaps we can therefore conclude that there does not exist a 
simple aligned magnetic field geometry in the core of Arp~220.
The extreme dust environment in Arp~220's core, however, makes it dangerous to
draw conclusions based on dust properties in more benign environments like
that in our own Galaxy or even M82.

\citet{jones89} detect weak near-IR polarization in Arp~220, which they
interpret as due to a simple screen of aligned dust in front of a bright
nucleus.  The reported value is $0.54\pm0.18$ per cent polarization at
$K$-band, with a position angle of $58^\circ\pm11^\circ$.
\citet{siebenmorgen01} report a mid-infrared detection of polarization due to 
absorption
of $3.1\pm0.9$ per cent at 14.3$\,\mu$m with a position angle of
$62^\circ\pm9^\circ$.
Both of these studies also conclude that the dust grain alignment must be
inefficient, otherwise the high optical depth would lead to a much higher
degree of polarization.  It is unclear, however,
if Arp~220's relatively low level of polarization is typical,
since \citet{siebenmorgen01} report mid-IR polarization levels as high as
8 per cent in other ULIRGs.

At different wavelengths one might expect the core polarization to reflect
the geometry of the nuclear disk, the separation of the double nucleus and
the orientation of outflows.  The gas
disc of Arp~220 is at approximately  ${\rm PA}=45^\circ$
\citep[e.g.][]{scoville97}, while
models of the core radio, mm and sub-mm emission suggest 2 components separated
by ${\sim}\,1$ arcsec, with ${\rm PA}\simeq80$--$100^\circ$
\citep[e.g.][]{baan95,scoville97,EckDow,MunFP}.
Obviously our upper limit sheds
little light on whether the magnetic field structure is related to any of
these features -- this will await more stable measurements, as well as higher
resolution studies.

A naive estimate for the level of polarized fluctuations from
extragalactic sources is the product of the level of fluctuations in flux
and the average percentage polarization \citep{dezotti99}. 
A more detailed calculation can be done using the source counts for
submillimetre galaxies.  The shot noise from these sources results in a
flat contribution to the CMB temperature or Stokes $I$ $C_\ell$s:
\begin{equation}
C^I_\ell = \int_0^{S_{\rm cut}} S^2 (dN/dS) dS,
\end{equation}
where $dN/dS$ are the differential source counts and $C_\ell$ is the angular power
spectrum.  The contribution to
the $E$- and $B$-mode power spectra are equal and given by
\begin{equation}
C^E_\ell = C^B_\ell = p^2 C^I_\ell/2,
\end{equation}
for sources having fractional polarization $p$ (see Tucci et al.~2004 and
references therein).
Assuming $p=1.5$ per cent and following the counts estimate
of Scott \& White (1999) we show in Fig.~2 the level expected if individual
sources can be removed at the $100\,$mJy level.  
We also show two estimates from de Zotti et al. (1999) at
143 GHz, assuming each galaxy is 2\% polarized.

In producing CMB angular power spectrum estimates, one can remove or marginalize over 
pixels in the map with clearly detected point sources. The contamination of 
concern is then generally due to the brightest sources immediately below the 
detection threshold.
The 100 mJy flux cut corresponds to the  
$4 \sigma$ detection limit for the \planck\ 353~GHz channel all sky survey
\citep{bluebook05}.
The \planck\ point source detection limits, however, can differ
depending on the assumptions about the foregrounds, the specific
method adopted, and whether data at multiple frequencies are used
\citep[e.g.,][]{vielva01,vielva03}.  We note that raising the flux cut level to 
200~mJy would result in roughly a 10 per cent
increase in the contribution to the angular power spectrum.

Also shown is an overall view of the level of expected CMB temperature and polarization fluctuations
using a standard cosmological model constrained by the WMAP results, constructed
using CMBFAST \citep{cmbfast}.
The $B$-mode fluctuation curve (the potential contribution from primordial gravitational waves) was constructed assuming a tensor-to-scalar ratio of 0.1. 

The \planck\ 143 GHz and 353 GHz pixel noise contribution to the 
polarized angular power spectrum sensitivity levels are also shown in Fig.~2. 
The standard error per $\ell$-mode is derived from the 
\planck\ instrument sensitivity \citep{bluebook05} using 
\begin{equation}
\sigma_{\ell}^P = \sqrt{\frac{2}{f_{\hbox{\small sky}}(2\ell + 1)}} (C_{\ell}^P + N_{\ell}^P ),
\end{equation}
where $\sigma_{\ell}^P$ is the standard error per mode in the polarization power spectrum,
$f_{\hbox{\small sky}}$ is the fraction of sky observed and $C_{\ell}^P$ is the angular power spectrum
of the CMB polarization signal
\citep[see, e.g.,][]{knox95,kesden02}. $N_{\ell}^P$ is the pixel noise contribution given by 
\begin{equation}
N_{\ell}^P = f_{\hbox{\small sky}} \frac{4 \pi s^2}{\tau} e^{ {\ell}^2 {\sigma_b}^2 },
\end{equation}
where $s$ is the effective sensitivity, $\tau$ is the total integration time, and $\sigma_b$
is the instrument beam gaussian width.
The noise estimates here are simple ones and do not account for the effects of non-uniform sky
coverage, galactic cuts, detector $1/f$ noise, etc.  Nevertheless,
the \planck\ noise levels are more than an order of 
magnitude above the polarized dusty galaxy level and we therefore conclude that 
such galaxies are unlikely to contaminate CMB polarization 
measurements with \planck\, even in the higher CMB frequency bands. 

Contamination in the 350 GHz range may be an issue, however, for future
CMB polarization measurements that attempt to reach the 0.01 tensor-to-scalar
ratio level, particularly at the higher multipoles. Experiments designed to
measure the lensing contribution to $B$-mode polarization may also need to 
carefully consider the potential impact of polarized extragalactic sources.

Our current state of knowledge, however, is still incomplete. The above analysis
relies on a number of extrapolations
and simplifying assumptions which may not be correct, including: the unknown
source counts at the ${\sim}\,100\,$mJy level; precisely at what level
individual sources can be removed; the real distribution in the
degree of polarization of such objects; and whether clustering, potentially
involving partial alignments of polarization axes, might be significant.
Determining whether Arp~220's level of sub-mm polarization is actually typical will await measurements of more 
objects and greater control of systematic
effects, which will be possible with the new SCUBA-2 instrument.

\section*{Acknowledgments}
It is a pleasure to thank Jane Greaves, Gerald Moriarty-Schieven, and
Barth Netterfield for
useful discussion. We also thank the entire JAC staff for their assistance.
The James Clerk Maxwell Telescope is operated by The Joint Astronomy Center on
behalf of the Particle Physics and Astronomy Research Council of the United
Kingdom, the Netherlands Organisation for Scientific Research, and the National
Research Council of Canada.  This work was
supported in part by the Natural Sciences and Engineering Research Council 
of Canada. The research described in this paper was performed in part at the 
Jet Propulsion Laboratory, California Institute of Technology, under a contract 
with the National Aeronautics and Space Administration.

\end{document}